\documentclass[prl,twocolumn,preprintnumbers]{revtex4}
\usepackage{graphicx}
\usepackage{dcolumn}
\usepackage{bm}
\usepackage{amsmath}
\usepackage{amsfonts}
\usepackage{amssymb}
\usepackage[latin1]{inputenc}

\begin{document}

\title{Three-dimensional chiral meta-atoms}
\author{Carsten Rockstuhl}
\author{Christoph Menzel}
\author{Thomas Paul}
\author{Falk Lederer}
\affiliation{Institute of Condensed Matter Theory and Solid State Optics
\\Friedrich-Schiller-Universität Jena, Max-Wien-Platz 1, D-07743 Jena,
 Germany}

\begin{abstract}
We show that the chirality of artificial media, made of a planar
periodic arrangement of three-dimensional metallic meta-atoms, can
be tailored. The meta-atoms support localized plasmon polaritons and
exhibit a chirality exceeding that of pseudo-planar chiral
metamaterials by an order of magnitude. Two design approaches are
investigated in detail. The first is the referential example for a
chiral structure, namely a Moebius strip. The second example is a
cut wire - split-ring resonator geometry that can be manufactured
with state-of-the-art nanofabrication technologies. Driven into
resonance these meta-atoms evoke a polarization rotation of
$30^\circ$ per unit cell.
\end{abstract}

\maketitle

Metamaterials are artificial structures usually composed of
periodically arranged unit cells that allow to control the
properties of light propagation\cite{NaderBook}. Frequently the
rigorous description of light propagation in such media on the basis
of the dispersion relation of the respective eigenmodes (Bloch
functions)\cite{Carsten2} can be simplified by treating the medium
as effectively homogenous\cite{Menzel}. The properties usually at
the focus of interest are permittivity and permeability or
refractive index and impedance. However, to obtain more complex
optical functionalities efforts have been undertaken to extend this
concept towards other optical properties, as e.g.
chirality\cite{PendryChiral}.

By definition, a structure is termed \textit{chiral} if the unit
cell can not be mapped onto its mirror image by proper rotations.
Consequently, only a bulk medium with three-dimensional unit cells
can exhibit this property. Chiral media attract much interest
because the optical response of these structures is different for
right- and left-handed circularly polarized light. Thus, these media
are optically active and the state of polarization of light changes
upon traversing such media. The observation of this phenomenon by
using appropriately structured metallic thin-films on substrates
sparked significant research interest on artificial chiral
media\cite{PRA_OptActTheory,PRL_OptManPlaChir,PRL_GiantZheludev,JOA_PottsBagnall}.
These structures were termed planar chiral metamaterials (PCM) as,
at a first glance, the unit cell is a thin film only with no
structural variation in the propagation direction. An ensemble of
gammadions is one preferential geometry of such a PCM. Although the
term PCM is an oxymoron, one usually argues that the presence of the
substrate breaks the mirror symmetry and saves the three dimensional
character of the unit cell\cite{WegenerChiral}. However, the optical
activity of these tiny pseudo-PCMs is small, leaving much space for
further studies. To date potentially the largest optical activity
was observed for a gammadion bilayer where the gammadions in
subsequent layers are rotated by $15^\circ$ with respect to each
other\cite{PlumAPL}. Although the rotation per unit cell of this
pseudo-PCM at the resonance wavelength was orders of magnitude
larger than that of any naturally available substance, it is still
rather small amounting to $0.37^\circ$. The question arises if there
are feasible approaches towards larger optical activity and which
enhancement can be achieved if the pseudo-PCMs are replaced by
appropriately designed 3D unit cells. It is the aim of this Letter
to disclose effective design principles and to investigate the
performance of a new class of 3D chiral meta-atoms.

Prior to further discussions we stress that the chirality explored
here is related to the symmetry of the unit cell rather than to an
appropriate rotation of adjacent crystal planes, as e.g. naturally
observed in quartz or in cholesteric liquid crystals. In the latter
structure, being potentially the optically most active naturally
available substances, optical activity can be as large as $3\times
10^{4}\,^{\circ}/mm$ \cite{Cholesteric}. Such structures can be also
mimicked by fabricated spiral type photonic
structures\cite{WegenerChiral_2,Yannopapas} or chiral sculptured
thin films (CSTF)\cite{LakhtakiaCSTF,LakhtakiaAmbichiral}. The final
aim will be to construct an effective chiral metamaterial from these
meta-atoms. Therefore, we furthermore exclude geometries that show
the desired optical response exclusively for a specific angle of
incidence, a specific incidence polarization state or even in a
higher diffraction order.

The design rules for the 3D chiral meta-atoms we suggest follow
simple physical principles. Large optical activity requires the
meta-atom to exhibit two resonances in two different elements of the
meta-atom. The first resonance should be excitable by the incoming
light providing sufficiently strong coupling between the external
field and the meta-atom. The second resonance should appear for
orthogonally polarized light in another element of the meta-atom
serving to generate a radiated field that is orthogonally polarized
with respect to the incident field. Both structural elements have to
be sufficiently coupled to allow for an efficient polarization
conversion. The absence of any mirror symmetry in the unit cell
ensures that coupling between the two modes sustained by the
structure is not prohibited due to symmetry constraints. Within the
quasi-static approximation this leads to the conclusion that the
meta-atom has to be composed of two resonant structural elements
mimicking a dipole type scattering response. These two elements can
be either two electric dipoles or an electric and a magnetic dipole.
A structural element having an electric dipole resonance is e.g. a
metallic wire of finite length\cite{Kreibig} whereas a split-ring
resonator (SRR)\cite{origin} exhibits a magnetic dipole resonance.

We will study meta-atoms that follow either design principles. This
approach towards the understanding of optical activity of meta-atoms
provides an extremely intuitive explanation of the fundamental
physical effect, as will be shown below. In this context it might be
interesting to apply this approach also in other fields, as e.g. in
chemistry, for understanding chirality and optical activity of
various naturally available molecules and macro-molecular
structures\cite{JACS_ContinousSymmetryMeasures}. The meta-atoms we
consider have a 3D geometry. Building a genuine 3D material they can
be periodically arranged in succeeding the $x - y$-planes without
any twist between adjacent planes. Thus optical activity based on
that twist is safely excluded. Because the meta-atom does not
exhibit any mirror-symmetry optical activity is expected to occur
for any propagation direction and its strength is only determined by
the particular geometrical arrangement within unit cell. Without
loss of generality we assume the light to propagate in
$z$-direction.

The asymmetry of the unit cell requires a general bi-anisotropic
description of the effective medium composed of the
meta-atoms\cite{TretyakovBook}. Hence, the rotation of the
polarization is a consequence of the joint action of optical
activity and birefringence of the structure. Both effects may be
discriminated, if the polarization rotation is averaged over all
possible linear polarization states of the incident light. The
averaged rotation is then a measure of optical activity and
chirality of the meta-atoms. This is intuitively clear if a
statistical ensemble of meta-atoms, arbitrarily rotated around the
$z$-axis, is considered where a light field experiences a
bi-isotropic rather than a bi-anisotropic medium.

\begin{figure}[h]
\centering
\includegraphics[width=85mm,angle=0] {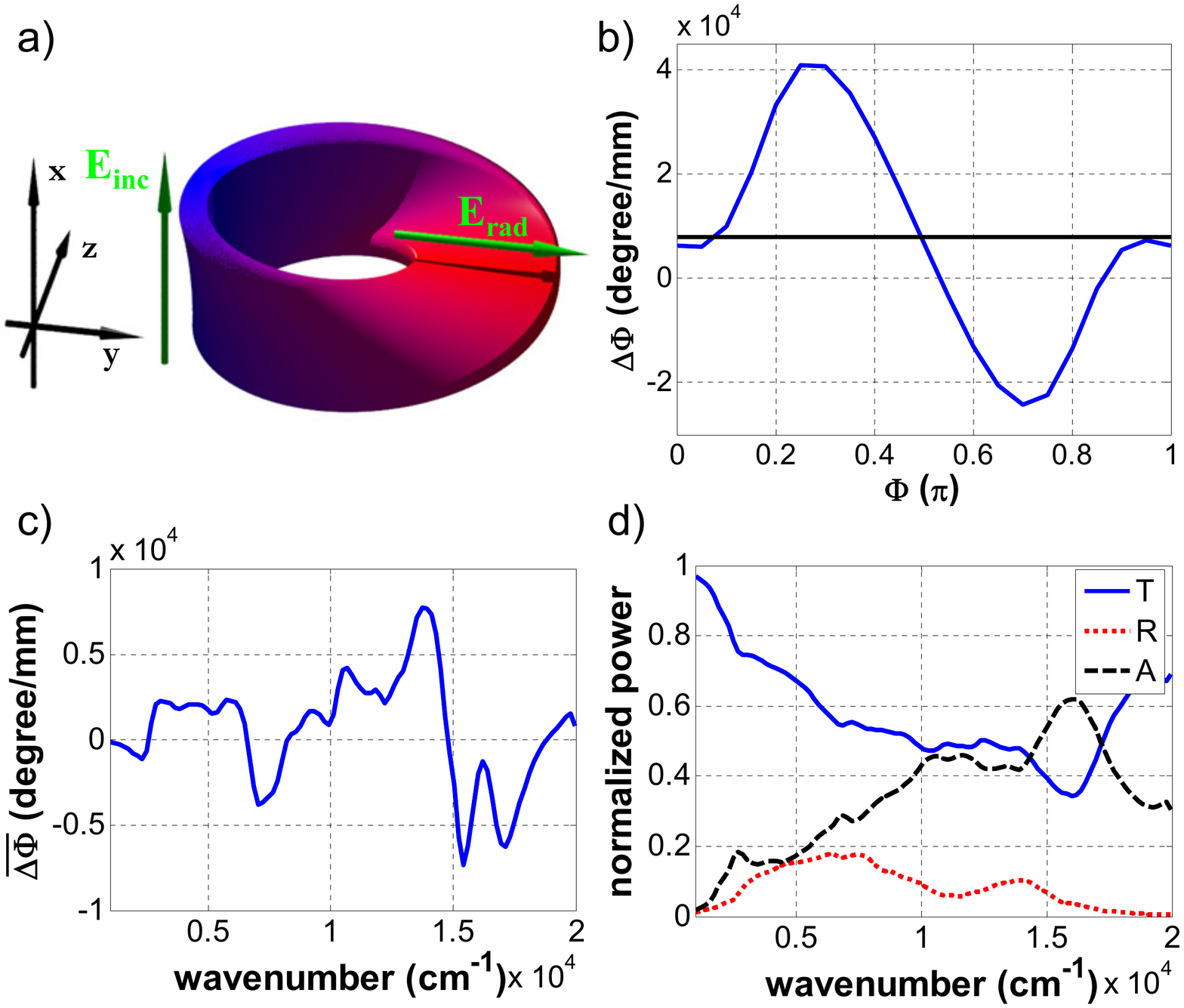}
\caption[Submanifold]{(color online) The Moebius strip. a) Geometry and coordinate
system; the arrows indicate the polarization of the electric dipole resonances. b)
Polarization rotation as a function of the polarization angle of the incident field at
$\nu=1.373\times 10^{4}\,cm^{-1}$. c) Averaged polarization rotation as a function of the
wavenumber. d) Averaged reflection, transmission, and absorption as a function of the
wavenumber.} \label{FIG_MOEBIUS_DATA}
\end{figure}

\begin{figure}[h]
\centering
\includegraphics[width=85mm,angle=0] {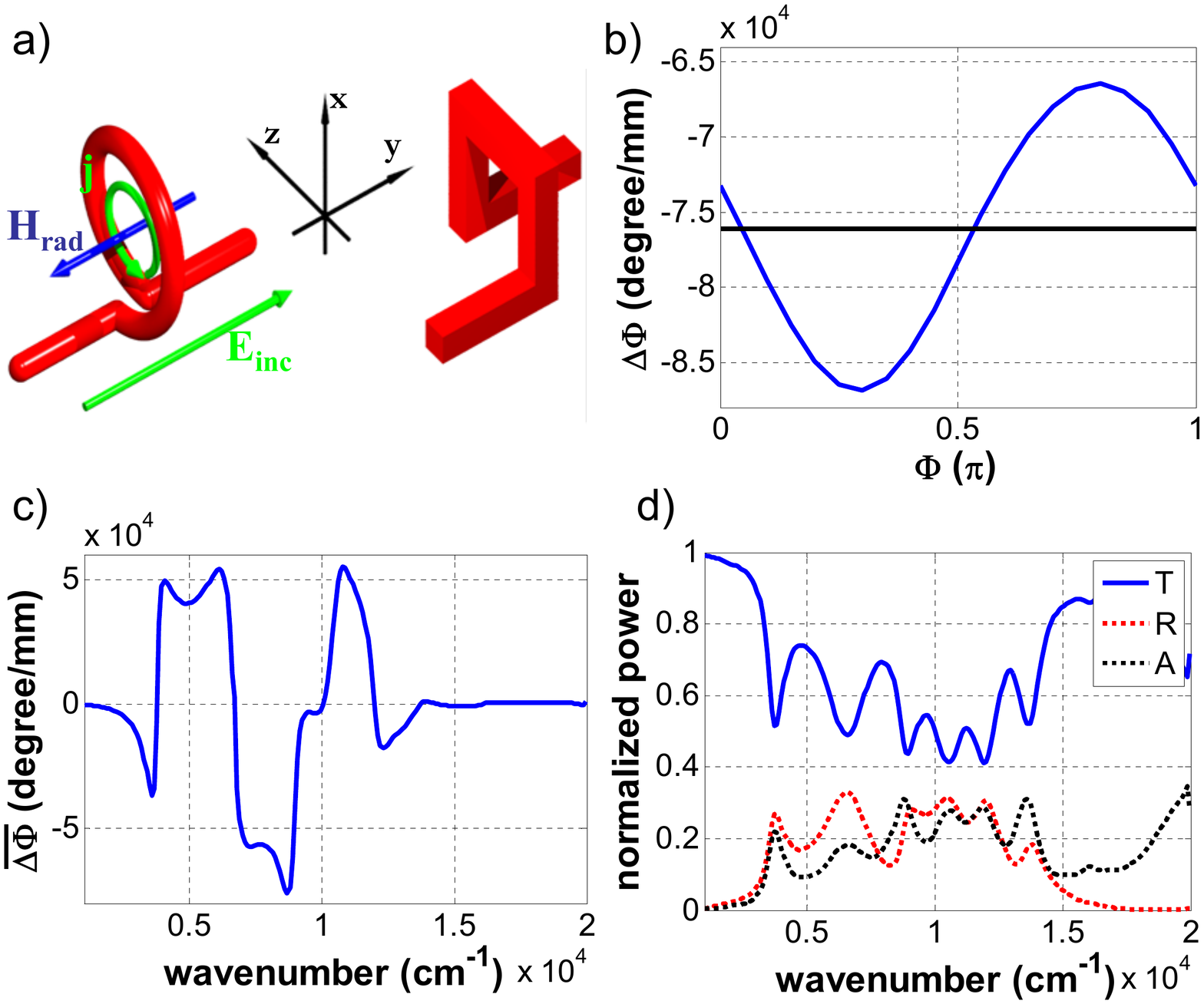}
\caption[Submanifold]{(color online)  The Omega particle. a) Geometry and coordinate
system; left: modified omega particle, right: cut wire-SRR particle used in numerical
simulations; the arrows indicate the polarization of the fields associated with the
electric and magnetic dipoles. b) Polarization rotation as a function of the polarization
angle of the incident field at $\nu=0.87\times 10^{4}\,cm^{-1}$. c) Averaged polarization
rotation as a function of the wavenumber. d) Averaged reflection, transmission, and
absorption as a function of the wavenumber.} \label{FIG_OMEGA_DATA}
\end{figure}

To validate our concept we study light interaction with a planar,
periodic arrangement of two different 3D meta-atoms. The optical
activity of the first, a Moebius strip \cite{MoebiusRef}( see
Fig.~\ref{FIG_MOEBIUS_DATA}(a)), relies on the interaction of two
resonant electrical dipoles, and the second, a rationally designed
cut wire - split-ring geometry (see Fig.~\ref{FIG_OMEGA_DATA}(a)),
takes advantage of the resonant interaction of an electric with a
magnetic dipole.

The terminating surface of the Moebius strip is characterized by the parametric equations
\begin{eqnarray}
x(u,v)&=&\frac{v}{2}\sin \left(\frac{u}{2}\right)\\
y(u,v)&=&\left[1+\frac{v}{2}\cos\left(\frac{u}{2}\right)\right]\cos
\left(u\right)\nonumber\\
z(u,v)&=&\left[1+\frac{v}{2}\cos\left(\frac{u}{2}\right)\right]\sin
\left(u\right)\nonumber,
\end{eqnarray}
with $0~\leq~u~\leq~2\pi$ being an angle and $-1~\leq~v~\leq~1$ being the normalized
radius and the width of the strip. In the numerical analysis we assumed a gold strip (
dielectric function from \cite{Christy}), where the radius and the width are identical
(166 nm) and the thickness amounts to 33 nm, surrounded by air. Evidently, any other
dielectric environment will only cause quantitative changes. The Moebius strips are
periodically arranged in the $x-y$ - plane with a period of 500 nm
in both directions. The chosen geometry ensures that the meta-atoms
are significantly sub-wavelength in the spectral domain of interest.
Only a zero order reflected and transmitted wave will be observed.
The orientation of the strip with respect to the axis as well as the
exact geometrical parameters are chosen arbitrarily to a certain
extent and shall only serve to illustrate the described approach. We
do not intend to optimize the chiral response. Nonetheless, the
particular orientation was chosen because certain segment of the
strip resemble cut wires in both the $x$ - and $y$ - direction. As
both wires are orthogonally oriented we expect that both segments
support the excitation of a localized plasmon polariton with an
electric dipole field at a certain frequency, though they require an
orthogonal polarization for their excitation. The excited electric
dipole and the radiating electric dipole are indicated in
Fig.~\ref{FIG_MOEBIUS_DATA}(a) by arrows.

Since the system is linear, it suffices to calculate the
transmission coefficients for two mutually orthogonal incident
polarizations, say $x$ and $y$ polarization. The transmission
coefficients $T_x$ and $T_y$ for arbitrary linear polarization
characterized by the polarization angle $\phi\in[0,2\pi)$ of the
incident light are given by
\begin{equation}
\begin{pmatrix} T_x\\T_y\end{pmatrix}=\hat{T}\begin{pmatrix}
I_x\\I_y\end{pmatrix}=\begin{pmatrix} T_{xx} & T_{yx}\\T_{xy} & T_{yy}\end{pmatrix}\begin{pmatrix} \cos(\phi)\\
\sin(\phi)\end{pmatrix}\label{EQ_T_MATRIX}
\end{equation}
where the matrix $\hat{T}$ is obtained by the two calculations mentioned before.

The average polarization rotation $\overline{\Delta\Phi}$ is then given by
\begin{equation}
\overline{\Delta\Phi}=\frac{1}{2\pi}\int_{0}^{2\pi}\Delta\Phi(\phi)\,d\phi\label{EQ_DELTA_PHI_QUER}
\end{equation}
where the polarization rotation $\Delta\Phi(\phi)$ is given by
\cite{APL_Menzel}
\begin{equation}
\Delta\Phi(\phi)=\Re\left\{\textrm{atan}\left(\frac{T_\bot(\phi)}{T_\|(\phi)}\right)\right\}
\label{EQ_DELTA_PHI}
\end{equation}
with $T_\|$ denoting the transmitted amplitude parallel polarized
and $T_\bot$ denoting the transmitted amplitude perpendicular
polarized to the incident field.

The complex transmission coefficients have been computed by using
the Fourier Modal Method (FMM)\cite{Li1}.

For a wavenumber of $\nu=1.373\times 10^{4}\,cm^{-1}$ the
polarization rotation as a function of the polarization angle of the
incident light is shown in Fig.~\ref{FIG_MOEBIUS_DATA}(b). As usual
in the field of chiral metamaterials, the polarization rotation is
measured in angle per mm. The average value of the polarization
rotation is indicated by a straight line and amounts to
$\overline{\Delta\Phi}=7.711\times 10^{3}\,^{\circ}/mm$; or
$3.5^{\circ}$ per meta-atom layer. This polarization rotation is
evoked by the chirality of the metamaterial and compares to that of
common PCMs. The averaged polarization rotation as a function of the
wavenumber is shown in Fig.~\ref{FIG_MOEBIUS_DATA}(c). Because the
structure meets always the symmetry requirements (two orthogonal cut
wires), optical activity is present in the entire spectral domain.
The strength depends on the respective resonance strength, which
depends critically on the wavenumber. A pronounced maximum exists
around $\nu=1.373\times 10^{4}\,cm^{-1}$. The angular averaged and
spectral dependent optical coefficients (transmission, reflection,
and absorption) are shown in Fig.~\ref{FIG_MOEBIUS_DATA}(d).
Spectral regions of high absorption correspond to spectral regions
of high optical activity. This backs our initial statement that
strong optical activity is linked to the excitation of a localized
plasmon polariton resonance. Nonetheless, it remains noteworthy that
even in the spectral domain of the strongest absorption the
transmission is as large as about 40 percent.

The second design approach for a chiral meta-atom uses the
combination of structural unit elements that support the excitation
of an electric dipole and a magnetic dipole resonance. The former is
again a finite metallic wire element, whereas the latter is provided
by a SRR. The idealized geometry we envision is shown in the left of
Fig.~\ref{FIG_OMEGA_DATA}(a).
To a certain extent the structure resembles an omega particle,
though the metallic wires are perpendicularly arranged with respect
to the SRR plane. This structure was already studied in the context
of metamaterials but only with regard to controlling the
permittivity rather than its optical activity \cite{Smith99}. The
interaction of the light with the meta-atom can again be best
understood in terms of elementary electromagnetic excitations. An
$y$-polarized incident electric field induces at first an
oscillating current in the wire. At the plasmon polariton resonance
the charge density oscillation is in resonance. As the SRR is
coupled to the wire, a current will also flow through the wire
forming the SRR. The oscillating current will induce a magnetic
field that is perpendicular to the SRR from which it will radiate.
The radiated magnetic dipole has a dominant electric field component
perpendicular to the incident polarization. This field will cause a
strong rotation of the polarization, hence maximizing optical
activity of the meta-atom.

Because the fabrication of such a highly curved structure would be
seemingly difficult at optical frequencies, we suggest and study a
slightly simplified variant (right section of
Fig.~\ref{FIG_OMEGA_DATA}(a)) that maintains, however, all
structural features. Basically all curved elements were replaced by
their rectangular counterparts. In the subsequent numerical analysis
the cross-section of the wire amounts to 50 nm $\times$ 50 nm. The
length of the nanowires is 200 nm and the length of the SRR side
arms as well as of the SRR base is 300 nm. The SRR is again made of
gold and the period in both the $x$ - and $y$-direction is 500 nm.

The polarization rotation as a function of the polarization angle of
the incident field at a wavenumber of $\nu=0.87\times
10^{4}\,cm^{-1}$ is shown in Fig.~\ref{FIG_OMEGA_DATA}(b). Two
features are important to note. Optical activity is larger by an
order of magnitude when compared to the Moebius strip and the
structure exhibits only a marginal anisotropy. The averaged
polarization rotation is large and varies only slightly (about 1/5
of the averaged rotation) with the polarization angle of the
incident field. Note that the variation for the Moebius strip was
about 4 times the averaged value. The average polarization rotation
at the pertinent wavenumber amounts to
$\overline{\Delta\Phi}=7.609\times 10^{4}\,^{\circ}/mm$; or
$30.43^{\circ}$ per meta-atom layer. The spectrally dependent
average polarization rotation is shown in
Fig.~\ref{FIG_OMEGA_DATA}(c). It shows well pronounced resonances
where it is significantly enhanced within spectrally narrow domains.
The position of the resonances can be unambiguously correlated to
resonances in the absorption or the transmission, shown in
Fig.~\ref{FIG_OMEGA_DATA}(d). It is likewise averaged over all
possible linear polarization states of the incident field. Near all
resonances steep changes occur for the polarization rotation. There,
either the cut wires or the SRRs or both support the excitation of a
localized plasmon polariton. However, it is pointless to study in
detail the isolated resonances. One should rather regard the
meta-atom as a single complex structure that allows the excitation
of localized plasmon polaritons with various distinctive
polarization components. The strength of optical activity depends
then on the coupling of the incident field to this eigenmode.

In conclusion, we have systematically analyzed the optical activity
of planar periodic arrangements of three dimensional chiral
meta-atoms. The working principle of these meta-atoms was explained
in terms of elementary excitations of the structure. A large optical
activity was predicted to occur if the meta-atom contains two
structural elements, which may exhibit polariton plasmon resonances
with orthogonal polarizations. If these structural elements are much
less than the wavelength of interest, the excitations represent
either electric or a magnetic dipoles. These considerations permit
the rational design of 3D meta-atoms forming the building blocks for
a 3D chiral metamaterial. Chiral meta-atoms relying on the coupling
of an electric with a magnetic dipole (a cut wire - SRR geometry)
proved superior to a meta-atom with two coupled electric dipoles.
The optical activity for a planar arrangement of the former
structure was as large as $30.43^{\circ}$ per layer. In this respect
these meta-atoms exhibit a giant chirality. Changing the geometry of
the essential structural units permits spectral the tuning of the
chiral properties. Based on this idea this work hosts a methodical
aspect as well that allows both to derive rational design strategies
for chiral unit cells and to disclose the basic physics of the
working principle of highly efficient chiral media.

\end{document}